**Boundary Effects in Biological Planar Networks: Pentagons Dominate Marginal Cells**

Kai Xu[1,*] and Fei He[2]

[1] Fisheries College, Jimei University, Xiamen, 361021, China

[2] School of Mathematical Sciences, Xiamen University, Xiamen, 361005, China

Contact author: kaixu@jmu.edu.cn, kxu2013@gmail.com

**Abstract** The topological and geometrical features at the boundary zone of planar polygonal networks remain poorly understood. Based on observations and mathematical proofs, we propose that marginal cells in *Pyropia haitanensis* thalli, a two-dimensional (2D) biological polygonal network, have an average edge number of exactly five. We demonstrate that this number is maintained by specific division patterns. Furthermore, we reveal significant limitations of Lewis law and Aboav-Weaire law by comparing the topological and geometrical parameters of marginal cells and inner cells. We find strong boundary effects that are manifested in the distinct distributions of interior angles and edge lengths in marginal cells. Similar to inner cells, cell division tend to occur in marginal cells with large sizes. Our findings suggest that inner cells should be strictly defined based on their positional relationship to the marginal cells.

## 1. Introduction

The topological and geometrical features of two-dimensional (2D) trivalent polygonal networks have been extensively studied for several decades [1]. Analyzing the evolutionary dynamics of 2D polygonal networks not only helps uncover the fundamental laws of nature but also provides new insights into the biophysical and biomathematical mechanisms underlying life phenomena. Two empirical laws, Lewis law and Aboav-Weaire law, have been widely used to describe the relationships between cell size, edge number, and spatial arrangement in 2D planar tiling [1, 2]. These laws have been validated across various 2D polygonal networks, ranging from atomic to cosmic scales [3, 4] and covering plants, algae and animals [5-8].

To our knowledge, previous studies have primarily focused on the inner regions of such networks, where all polygonal cells are fully surrounded by neighboring cells. However, certain 2D biological networks, such as the thallus of *P. haitanensis*—comprising of a monolayer of polyhedral cells—exhibit a smooth boundary. In these networks, one edge of each marginal cell is not shared with any other cell, and the coordination number of every vertex is strictly equal to three. Intuitively, the marginal cells display distinct topological and geometrical features compared to inner cells.

To address this gap, we quantified the edge number, size, edge length, interior angle and neighboring relationship of marginal cells in *P. haitanensis*. Additionally, we analyzed the relationship between cell division and cell shape parameters. Our findings provide valuable insights that may be applicable to a wide range of scientific and practical contexts.

## 2. Methods

The software Amscope Toupview 3.0 was used to analyze the topological and geometrical parameters of cells from the *P. haitanensis* strain Sansha, as well as red and green mutant cells from the strain WO. The images of *P. haitanensis* tissues were obtained using a Zeiss Axioscope 5 microscope (Carl Zeiss, Germany) equipped with a Axiocam 208 color camera. In this study (Fig 1A), marginal cells (MCs) were defined as those located at the boundary of the polygonal cell network. Neighboring cells of MCs (NC-MCs) were characterized by their adjacency to MCs, sharing at least one common edge. Inner cells (ICs) are fully enclosed within the network, with no direct connection to NC-MCs. Each marginal cell contained a marginal edge (ME), positioned at the tissue boundary and belonging exclusively to the marginal cell. Additionally, each marginal cell contains two marginal angles (MA), which are directly associated with the ME (Fig 1A).

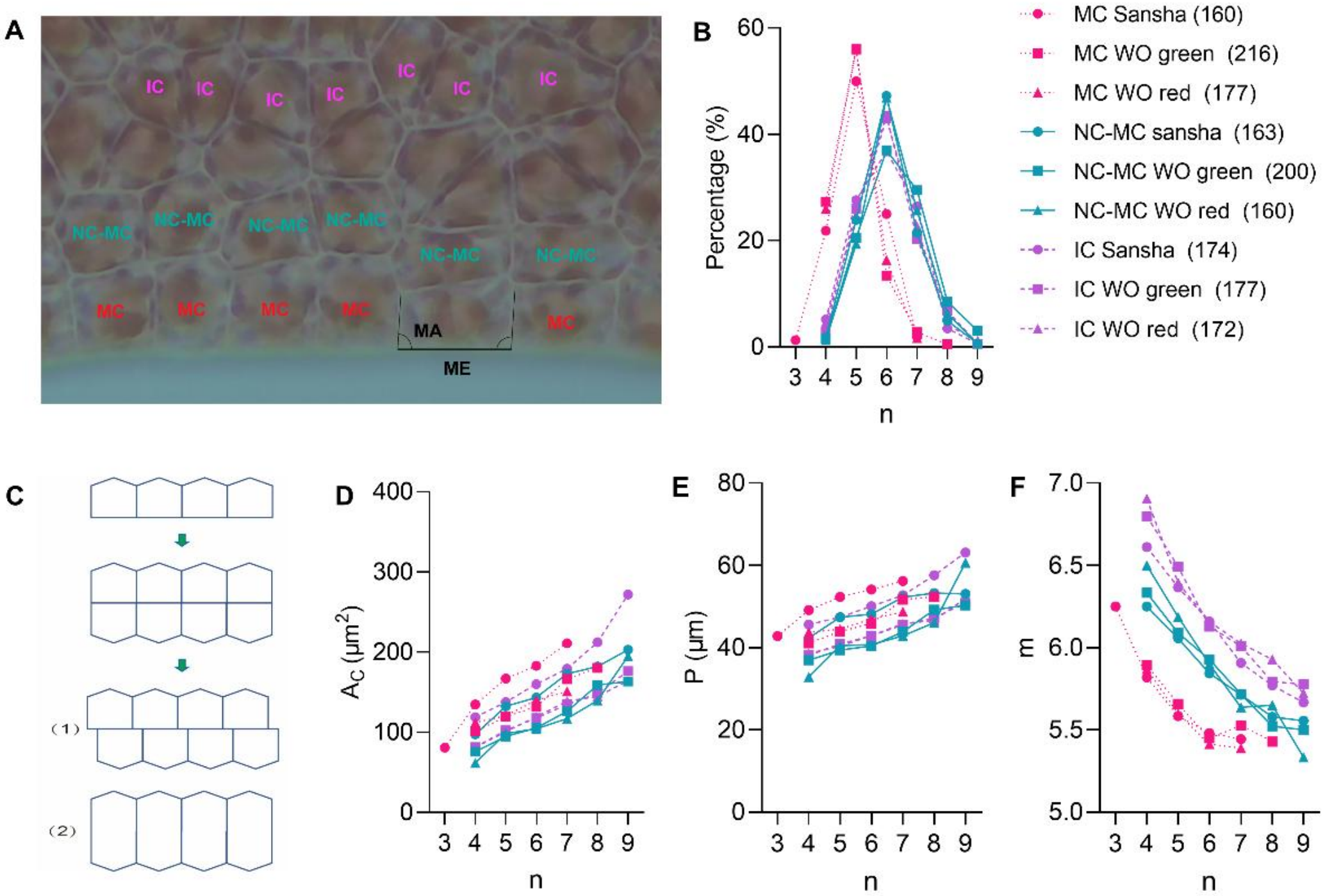

Figure 1 Topological and geometrical properties of *P. haitanensis* cells. (A) Positions of marginal cells (MCs), neighboring cells of MCs (NC-MCs), inner cells (ICs), marginal angles (MAs), and marginal edges (MEs) in *P. haitanensis* tissue. (B) Distribution of edge number $n$. (C) Diagram shows the methods of proof for $\bar{n}_{MCs} = 5$. Relationships between $n$ and cell area $A_C$ (D), $n$ and perimeter $P$ (E), $n$ and average edge number of neighboring cells $m$ (F). Numbers of examined cells are shown in parentheses.

For each polygonal cell, we counted the number of edges ($n$), and measured the area ($A_C$), perimeter ($P$), edge length, interior angles, and the coordinates of the center ($X_{PC}$, $Y_{PC}$) and vertices ($X_V$, $Y_V$). We used software R (version 4.0.0) with package Conicfit (version 1.0.4) to fit an ellipse based on the coordinates of the vertices of each polygonal cell [9]. The area of the ellipse's maximal inscribed polygon ($A_{EMIP}$) is given by $A_{EMIP} = 0.5nabsin(2\pi/n)$, where $n$ is the number of edges of inscribed polygon, $a$ is the semi-major axis, $b$ is the semi-minor axis [10].

The effects of cell division on topological and geometrical parameters were identified through time-series observations at time zero and 24 hours later. The size ratio of daughter cells was calculated as $SR = A_S \div A_L$, where $A_S$ and $A_L$ are the areas of the smaller and larger daughter cells, respectively. To quantify the relative position of the paired edges in the mother cell that are transected by cell divisions, the number of interval edges between the paired sides was counted in opposite directions, and the smaller number was used in this study [11].

## 3. Results and discussion

### 3.1. Edge number distribution

Lewis first reported two significant and conserved features of the edge number distribution in trivalent 2D polygonal networks: the average edge number $\bar{n}$ is six and hexagons are predominant [8, 12]. Here, we found that inner cells (ICs) have an average edge number $\bar{n}_{ICs}$ of approximately six (Tab S1) and are predominantly hexagons (Fig 1B). In contrast, marginal cells (MCs) have an average edge number $\bar{n}_{MCs}$ of approximately five and are predominantly pentagons. For inner cells, $\bar{n}_{ICs} = 6$ can be derived from Euler's 2D equation under two conditions: every three edges meet at a vertex, and the number of polygonal cells is sufficiently large [1]. Based on our results, we propose that previous studies reporting $\bar{n} = 6$ for trivalent 2D networks may have focused exclusively on ICs. These findings are consistent with our earlier study [11], which found that $\bar{n}$ increases from four to six as the *P. haitanensis* thalli proliferated from an initial group of 4-10 linearly ordered cells to a 2D tiling comprising a large number of cells. We suggest that the variation in $\bar{n}$ is closely associated with the proportion of MCs. Furthermore, $\bar{n}_{MCs} = 5$ for the marginal cells could be considered as a specific case of the inner cells. Below we provide two simple mathematical proofs (Fig 1C).

First method: Suppose the cells are very small relative to the boundary of the polygonal network, we can assume that the boundary is locally a straight line. By joining the polygonal network with its mirror image across the boundary and translating the mirror image a little to ensure that the coordination number equals three. In this new construction, each marginal cell can be considered as an inner cell. Suppose the edge number of a marginal cell is $n_{MC}$, then after translation the new cell has an edge number of $n_{MC} + 1$. Thus, on average sense, we should have $\bar{n}_{MCs} + 1 = \bar{n}_{ICs} = 6$, hence $\bar{n}_{MCs} = 5$. Second method: We assume that the edges intersect the boundary perpendicularly. By joining the polygonal network with its mirror image across the boundary and then erasing the boundary edges, then each marginal cell is doubled into an inner cell. After doubling the new cell has an edge number of $2(n_{MC}-1)-2$. Thus, we have $2(\bar{n}_{MCs}-1)-2=\bar{n}_{ICs}=6$, hence $\bar{n}_{MCs}=5$.

3.2. Lewis law and Aboav-Weaire law

Lewis also observed that the area of an $n$-edged cell increases linearly with $n$ [8]. This relationship, known as Lewis law, is considered a conserved (though empirical) law in trivalent 2D networks [1]. In this study, for each strain or mutant, the area $A_C$ of all three kinds of cells increased with $n$ (Fig 1D). However, the average values of $A_C$ of MCs, NC-MCs, and ICs were very similar (Tab S1). A similar trend was observed for cell perimeter $P$ (Fig 1E, Tab S1). These results suggest that cells of different sizes tend to distribute uniformly within the 2D polygonal network. Given the significant difference in $\bar{n}$ between MCs and ICs, Lewis law appears to hold locally but not globally.

The Aboav-Weaire law, another empirical law, states that the many-edged cell tends to neighbor with few-edged cells, and vice versa [13, 14]. In this study, the average edge number ($m$) of neighboring cells for all three cell types decreased as $n$ increased (Fig 1F). Additionally, according to the Aboav-Weaire law, $m$ for NC-MCs should decrease due to their adjacency to MCs. This hypothesis was confirmed, as the average $m$ for NC-MCs was generally lower than that for ICs (Tab S1). Due to the lower average $m$ of MCs and NC-MCs compared to that of ICs, the Aboav-Weaire law also appears to hold only locally. Specifically, Weaire's sum rule [14] is not applicable to MCs.

3.3. Ellipse packing

Previous studies suggests that polygonal cells in trivalent 2D networks tend to form the ellipse's maximal inscribed polygon (EMIP), a phenomenon termed ellipse packing [7, 11, 15]. This study found that the area $A_C$ of all cell types is approximately 0.85 times the area of the EMIP ($A_{EMIP}$) (Tab S1). This suggests that maintaining an optimal cell size may be more important for the cells (including MCs) than adhering to Lewis law.

Due to the restrictions of ellipse packing and $\bar{n}_{ICs}=6$, the interior angles of ICs peaked in the range of 110–130° (Fig 2A). Theoretically, $\bar{n}_{MCs}=5$ should cause the angle distribution of MCs to peak around 108°. However, about 42% of interior angles of MCs fell within 80–100°, primarily due to the fact that about 80% of the marginal angles (MAs), which accounted for roughly 40% of the interior angles, were in the range of 80–100°. This specific distribution of MAs may play a key role in reconciling the angular conflict between MCs and NC-MCs. Specifically, since two MAs of a five-edged MC are approximately 90°, the average angle of the three remaining angles (RAs) could approach 120°. This hypothesis is supported by our results, which show that the distributions of RAs and interior angles of ICs were similar (Fig 2A).

In addition, the distribution of edge lengths showed obvious differences between MCs and ICs (Fig 2B), which was mainly due to their similar mean perimeters but different average edge numbers (Tab

S1). Similarly, the edge length distribution of MCs was influenced by the marginal edges (MEs), which accounting for about 20% of MC edges. These results indicate strong boundary effects on the distributions of angles and edge lengths in MCs.

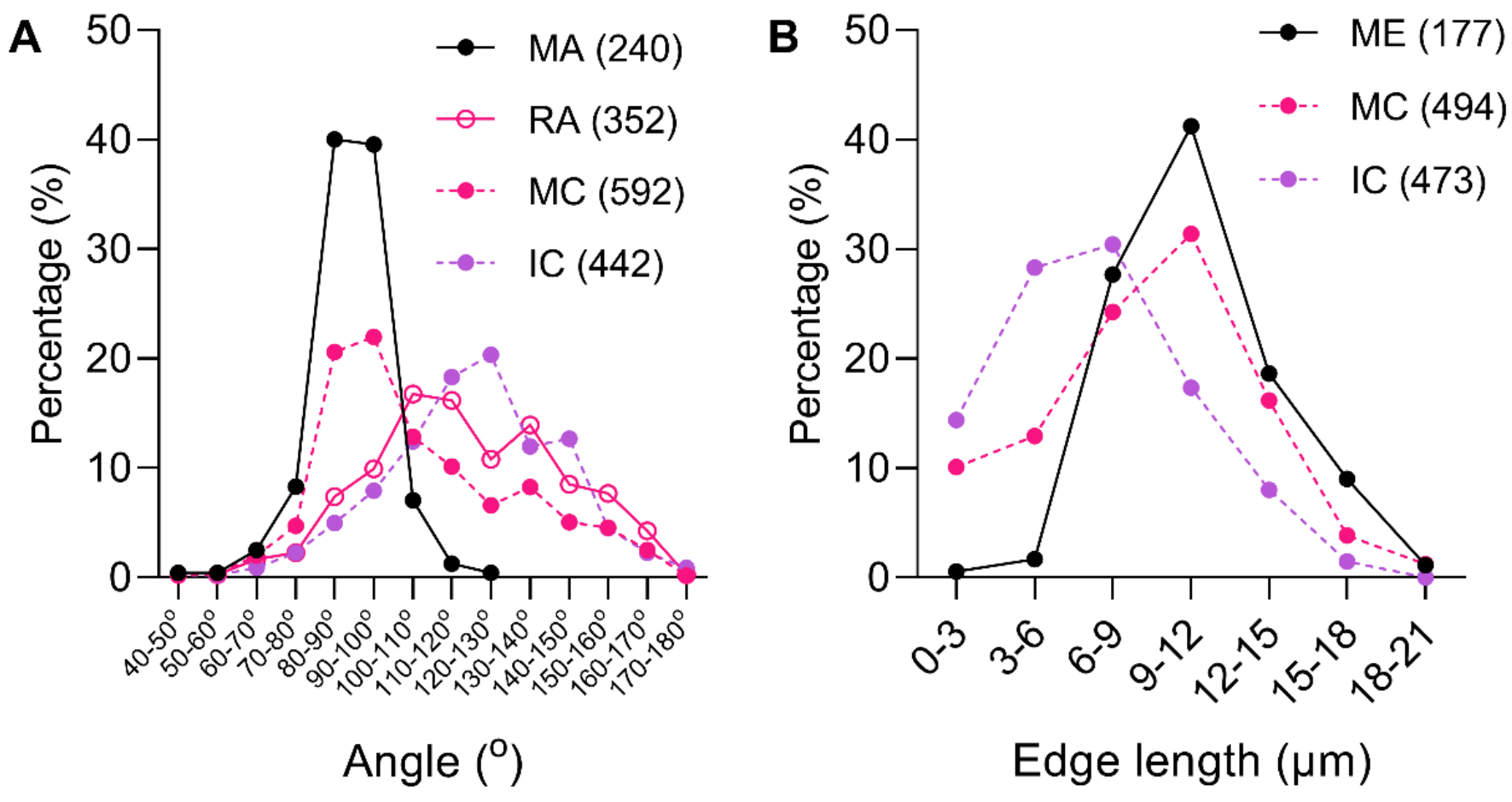

Figure 2 Angles and edge lengths of red mutant cells from strain WO. (A) Distribution of interior angles in MCs and ICs. On average, each MC contains two MAs and three remaining angles (RAs). (B) Distribution of edge lengths in MCs and ICs. Each MC contains only one ME. Sample numbers are indicated in parentheses.

### 3.4. Cell division

Previous studies suggest that mitotic division of *Drosophila* epithelia tends to occur in large cells which approximately contain one more edge compared with resting cells [16, 17]. Here, we found that, regardless of cell positions or strains (and/or mutants), the mean $A_C$ of dividing cells was at least 30% higher than that of resting cells (Fig 3A), while the mean $P$ of the dividing cells was 14~20% higher (Fig 3B). The mean $n$ of dividing cells was 6~10% higher than that of resting cells, while the differences in aspect ratio $a/b$ and $m$ were less than 5% (Fig 3C). These findings, along with previous studies [16, 17], suggest that cell size is a better indicator of dividing cells than $n$, $a/b$, and $m$.

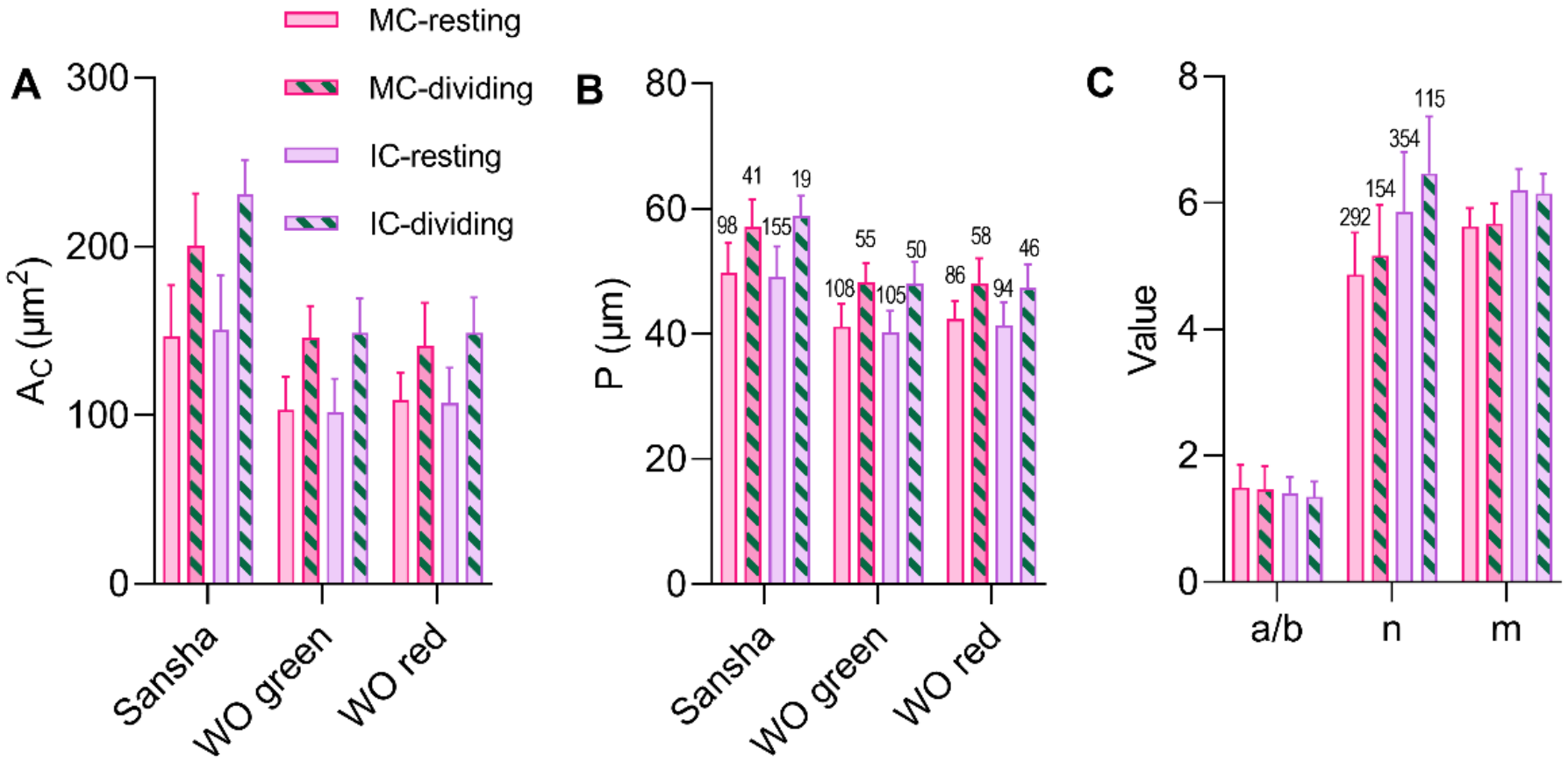

Figure 3 Differences between resting cells and dividing cells. (A) cell area $A_C$, (B) perimeter $P$, and (C) aspect ratio $a/b$, edge number $n$, and average edge number of neighboring cells $m$. For the latter three parameters, the data of all strains and mutants are combined. The numbers on bars denote the numbers of cells.

The division of an inner cell increases the cell number by one and the edge number by six, making $\bar{n}_{ICs} = 6$ a topological consequence of cell division in biological trivalent 2D networks [8, 12]. In general, mitosis tends to divide a cell equally, as verified by a previous study on ICs of *P. haitanensis* thalli [11] and confirmed in this study for MCs (Fig 4A). In addition, due to $\bar{n}_{MCs} = 5$, 96% of the number of interval edges were one (Fig 4B). Results from this study and a previous one [11] suggest that, for both ICs and MCs, cell division preferentially transects a pair of unconnected edges. The division of an MC that transects the ME produces two daughter MCs; otherwise, it produces only one daughter MC (Fig 4C). Based on the percentages of the five division types observed in MCs and the average number of dividing MCs ($\bar{n} = 5.16$) (Fig 3C), we found that the daughter MCs maintained an average edge number of 5.06. Therefore, $\bar{n}_{MCs} = 5$ is also a topological consequence of specific patterns of cell division. Further studies are needed to explore the underlying mechanisms governing these division patterns. Markov chain models have been used to further validate that the predominance of hexagons in epithelia is determined by cell division [18, 19]. The predominance of pentagons in marginal cells may also be explained by this model.

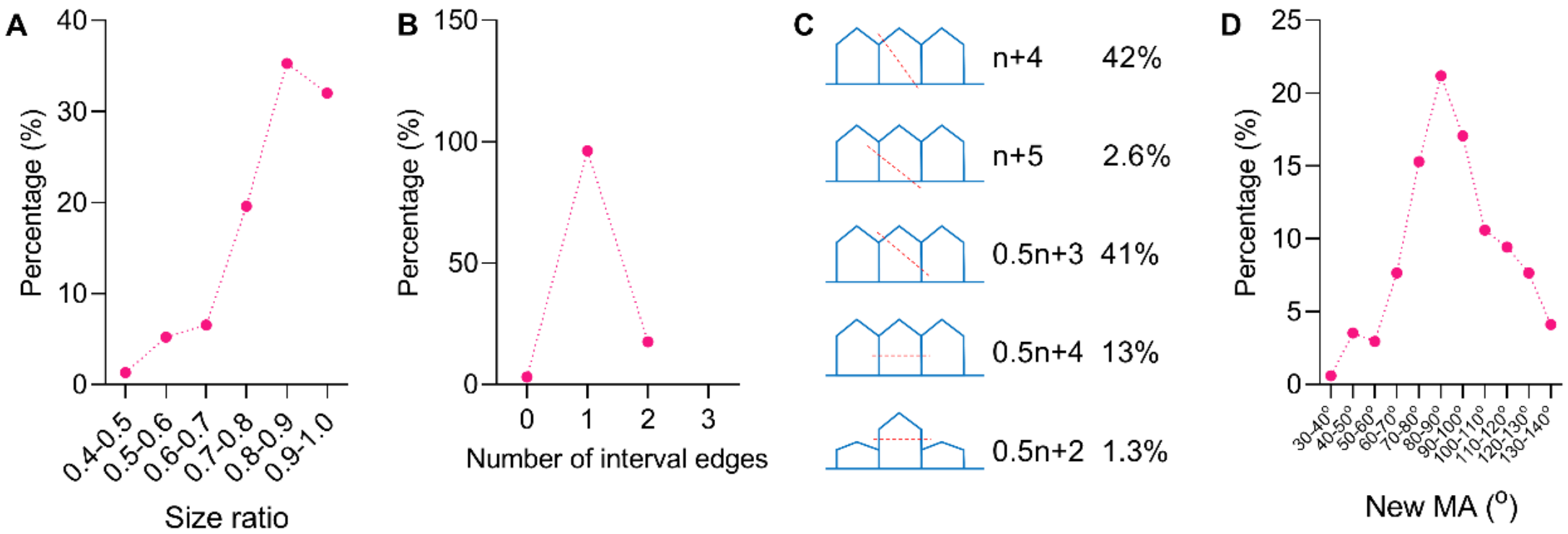

Figure 4 Cell division of MCs. (A) Size ratio of daughter cells. (B) Distribution of interval edges. (C) Five division patterns of $n$-edged MCs, showing how division alters both the number of MCs and their edge counts. Final daughter MC edge numbers (incorporating changes in neighboring MCs to simplify calculations) and the percentage of each division pattern are displayed to the right of panel (C). A total of 153 dividing MCs were examined. (D) Distribution of newly formed MAs.

In addition, while the specific preference of cell division has a profound effect on cellular geometries, post-division adjustments also play a significant role. For example, a previous study found that newly formed interior angles in ICs of *P. haitanensis* thalli rapidly adjusted to approach 120°[11]. We observed similar behavior in MAs in the present study. As shown in Fig 4D, the distribution of newly formed MAs differed significantly from the overall MA distribution (Fig 2A), indicating that newly formed MAs undergo subsequent adjustment toward 90°. This finding should be incorporated into simulation models to improve investigation of dynamic behavior in biological 2D networks. A pure mathematical modeling study revealed that ellipse packing drives the formation of the specific angle distribution of ICs and accounts for the manifestation of von Neumann-Mullins law governing cell area [15]. However, the effects of ellipse packing on MCs remain an open question.

**4. Conclusion**

We demonstrate that the average edge number $\bar{n}_{MCs}$ of marginal cells in *P. haitanensis* thalli is approximately five, which is maintained by specific division patterns. We also prove that $\bar{n}_{MCs} = 5$ for 2D polygonal network with a coordination number strictly equal to three. By comparing the topological and geometrical parameters of marginal cells and inner cells, we observed significant limitations of Lewis law and Aboav-Weaire law. In addition, the specific topological feature of marginal cells changed the distributions of interior angles and edge lengths. From a methodological perspective, our results suggest that inner cells should be strictly defined based on their positional relationship to marginal cells, and that inner cells and marginal cells should be analyzed separately regarding cell proliferation, Lewis law and Aboav-Weaire law.

**Acknowledgments**

We thank Lingchao Zhang for his assistance in seaweed culture and image acquisition. KX acknowledges support from the National Natural Science Foundation of China (Grant No. 42376109).

**Table S1** Parameters of polygonal cells and fitted ellipses in *P. haitanensis*. Parameters $n$, $m$, $A_C$, $P$, $a$, $b$, and $A_{EMIP}$ represent the edge number, average edge number of neighboring cells of an $n$-edged cell, area, perimeter, semi-major-axis, semi-minor-axis, and the area of the ellipse's maximal inscribed polygon, respectively.

| Parameters | MCs | | | NC-MCs | | | ICs | | |
|---|---|---|---|---|---|---|---|---|---|
| | Sansha | WO green | WO red | Sansha | WO green | WO red | Sansha | WO green | WO red |
| $n$ | 5.04±0.77 | 4.93±0.75 | 4.94±0.70 | 6.06±0.88 | 6.32±1.04 | 6.18±0.88 | 5.97±0.97 | 6.04±0.95 | 6.01±0.96 |
| $m$ | 5.62±0.31 | 5.69±0.31 | 5.64±0.29 | 5.86±0.29 | 5.86±0.34 | 5.88±0.32 | 6.16±0.32 | 6.19±0.34 | 6.21±0.33 |
| $A_C$ (μm$^2$) | 163.69±37.79 | 117.78±28.01 | 122.20±25.08 | 148.49±34.97 | 115.15±30.96 | 108.75±23.82 | 159.23±40.25 | 118.62±29.71 | 118.93±29.36 |
| $P$ (μm) | 52.02±5.59 | 43.67±4.87 | 44.76±4.36 | 49.02±5.84 | 42.21±5.46 | 41.59±4.68 | 50.13±5.68 | 43.10±4.98 | 43.00±4.76 |
| $a$ (μm) | 11.07±1.74 | 9.50±1.91 | 9.55±1.72 | 10.27±1.65 | 8.42±1.48 | 8.57±1.71 | 9.99±1.32 | 8.67±1.17 | 8.68±1.23 |
| $b$ (μm) | 7.66±1.11 | 6.40±0.98 | 6.70±1.02 | 6.84±0.94 | 6.06±0.85 | 5.99±0.86 | 7.28±0.94 | 6.28±0.89 | 6.35±0.82 |
| $a/b$ | 1.47±0.33 | 1.53±0.43 | 1.47±0.40 | 1.53±0.35 | 1.41±0.33 | 1.46±0.41 | 1.39±0.24 | 1.40±0.25 | 1.39±0.26 |
| $A_{EMIP}$ (μm$^2$) | 200.48±54.29 | 141.40±39.55 | 148.28±37.23 | 181.28±41.31 | 134.22±34.77 | 133.59±34.72 | 187.05±44.03 | 140.96±34.43 | 141.83±33.57 |
| $A_C/A_{EMIP}$ | 0.83±0.10 | 0.85±0.09 | 0.84±0.10 | 0.82±0.09 | 0.86±0.07 | 0.83±0.09 | 0.85±0.08 | 0.84±0.09 | 0.84±0.10 |